\documentclass[english]{article}

\usepackage{bm}  
\usepackage{graphicx}
\usepackage{amsbsy}
\usepackage{amsmath}
\usepackage{amsfonts}
\usepackage{paralist}
\usepackage{color}

\newcommand{\ebX}{\epsilon {\bf X}}
\newcommand{\bX}{\textbf{X}}

\author{G Gradenigo, A Puglisi, A Sarracino, D Villamaina, and A Vulpiani}
\title{Out-of-equilibrium generalized fluctuation-dissipation relations}

\begin{document}

\maketitle 
\tableofcontents

\pagebreak 

\section{Introduction}
\label{Introduction}

Surely one of the most important and general results of 
statistical mechanics is the existence of a relation between the
spontaneous fluctuations and the response to external fields of
physical observables. This result has applications both in equilibrium
statistical mechanics, where it is used to relate the correlation
functions to macroscopically measurable quantities such as specific
heats, susceptibilities and compressibilities, and in nonequilibrium
systems, where it offers the possibility of studying the response to
time-dependent external fields, by analyzing time-dependent
correlations.

The idea of relating the amplitude of the dissipation to that of the
fluctuations dates back to Einstein's work on Brownian motion~\cite{E05}. Later,
Onsager~\cite{O31,O31b} put forward his regression hypothesis stating that the
relaxation of a macroscopic nonequilibrium perturbation follows the
same laws which govern the dynamics of fluctuations in equilibrium
systems. This principle is the basis of the fluctuation-dissipation
relation (FDR) of Callen and Welton, and of Kubo's theory of
time-dependent correlation functions~\cite{K66}. This result represents a
fundamental tool in nonequilibrium statistical mechanics since it
allows one to predict the average response to external perturbations,
without applying any perturbation.

Although the FDR theory was originally obtained for Hamiltonian
systems near thermodynamic equilibrium, it has been realized that a
generalized FDR holds for a vast class of systems.  A renewed interest
toward the theory of fluctuations has been motivated by the study of
nonequilibrium processes. In 1993 Evans, Cohen and Morriss~\cite{ECM}
considered the fluctuations of the entropy production rate in a
shearing fluid, and proposed the so-called Fluctuation Relation (FR).
Such relation was then developed by Evans and
Searles~\cite{earlierpapers} and, in different conditions, by
Gallavotti and Cohen~\cite{GC}. Noteworthy applications of those
theories are the Crooks and Jarzynski relations~\cite{GK,CJ}.

In the recent years there has been a great interest in the study of
the relation between responses and correlation functions, also with the aim
to understand the features of aging an glassy systems.  While, on one hand, a
failure of the validity of the equilibrium statistical mechanics
(e.g. lack of the ergodicity) implies a violation of the fluctuation-
dissipation theorem,  on the other hand, as we will see in Sections
\ref{Comb model}, \ref{Langevin process}, \ref{GranularIntruder}, the
opposite is not true: a failure of the fluctuation- dissipation
theorem (at least in its ``naive version'') does not imply the
presence of glassy phenomena.  It may, for instance, resides in the
presence of stationary non-equilibrium currents.

\subsection{The relevance of fluctuations: few historical comments}
\label{The relevance of fluctuations: few historical comments}

Let us spend few words on the conceptual relevance of the fluctuations in
statistical physics.
Even Boltzmann and    Gibbs, who already knew the expression 
for the mean-square energy
   fluctuation, 
\begin{equation}
\langle(E-\langle E\rangle) ^2\rangle= k_B T^2 C_V 
\label{i.1}
\end{equation}
pointed out that fluctuations were too small
   to be observed in macroscopic systems:
{\it  In the molecular theory we
   assume that the laws of the phenomena found in nature do not
   essentially deviate from the limits that they would approach in the
   case of an infinite number of infinitely small molecules...~\cite{B96}
   ... [the fluctuations] would be in general vanishing quantities,
   since such experience would not be wide enough to embrace the more
   considerable divergences from the mean values}~\cite{G02}. 

On the contrary Einstein and Smoluchowski
attributed a central role to the fluctuations. 
   For instance Einstein understood very well that the well known equation~(\ref{i.1})
{\it ... would yield an exact
   determination of the universal constant }(i.e. the Avogadro number),
  {\it  if it were possible to determine the average of the square of the
   energy fluctuation of the system; this is however not possible
   according our present knowledge}~\cite{E04}.

It is well known that 
the study of the fluctuations in the Brownian motion
allows one for the determination
of the Avogadro number (i.e. a quantity at microscopic level) from
experimentally accessible macroscopic quantities.
Therefore we have a
non ambiguous link between the microscopic and macroscopic levels;
this is well clear from 
the  celebrated Einstein relation
which gives the diffusion coefficient $D$ in terms of macroscopic
variables and the Avogadro number:
\begin{equation}
  D= \lim_{t \to \infty}  {  {\langle (x(t)-x(0)^2 \rangle}
\over {2 t}}= \frac{RT}{6 N_{A} \pi \eta a} \,\, ,
\end{equation}
where $T $ and $\eta$ are  the temperature and  dynamic viscosity of the 
fluid respectively, $a$  the radius of the colloidal particle
$R=N_{A}/k_B$ is the perfect gas constant and $N_{A}$ is the Avogadro 
number.

    The theoretical work by Einstein~\cite{E05} and the experiments by Perrin~\cite{P13}
    gave a clear and conclusive evidence of the relationship between
    the diffusion coefficient (which is measurable at the macroscopic
    level) and the Avogadro number (which is related to the
    microscopic description). Therefore, already at equilibrium, small-scale
  fluctuations \emph{do matter} in the statistical description of a
  certain system. 

The main purpose of the present contribution is at
  showing \emph{how} such fluctuations must be accounted for when the
  system is  out of equilibrium. A relevant outcome will be the
  evidence that in order to properly describe small scale
  out-of-equilibrium fluctuations some \emph{correlations} not present
  at equilibrium must be taken into account.

The celebrated paper of Einstein, apart from its historical relevance
for the atomic hypothesis and the development of the modern theory of
stochastic processes, shows the first example of FDR.  Let us write
the Langevin~\cite{L08} equation for the colloidal particle of mass
$m$:
\begin{equation}
\label{BM.5}
{dV \over dt}=
 -\gamma V +\sqrt{ {{2 \gamma k_B T} \over m}  } \zeta \,\,
\end{equation}
where $\gamma=  6 N_{A} \pi \eta a/m$, and  $\zeta$ is a white noise, i.e. a 
Gaussian stochastic process
with $\langle\zeta(t)\rangle=0$ and $\langle\zeta(t)\zeta(t')\rangle=\delta(t-t')$.
An easy computation gives $D=\langle V^2\rangle\tau$.
Consider now a (small) perturbating force $f(t)=F\Theta(t)$,
where $\Theta(t)$ is the Heaviside step function.
It is simple to determine  the average  response of the
velocity  after a long time (i.e. the drift) to such a perturbation:
$\langle\delta V\rangle= F/ \gamma$.
Defining the mobility $\mu$ as:
$\langle\delta V\rangle= \mu F \,\,$
one easily obtains the celebrated Einstein relation (EFDR)
\begin{equation}
\mu={D \over {k_BT}} \label{EinsteinRelation}
\end{equation}
which gives a link between the diffusion coefficient (a property of
the unperturbed system) and the mobility which measures how the system
reacts to a small perturbation.  

An intersting point addressed here is that
  out-of-equilibrium conditions can be hardly obtained from a single
  Langevin equation when the noise and drag terms are local in
  time. Preserving this locality, and hence the Markovian nature of
  the model, we will see that out of equilibrium a proper description
  of the relation between correlations and responses can be given within  an enriched space of
  variables, where also some \emph{local} field coupled to the variable of
  interest must be considered.  This new field is itself a
  \emph{fluctuating} object whose dynamics is ruled by another
  stochastic equation.

\section{Generalized Fluctuation-Dissipation relations}
\label{Generalized Fluctuation-Dissipation relations}

 The Fluctuation-Response theory was originally developed in the context of
equilibrium statistical mechanics of Hamiltonian systems.
Sometimes this induced confusion, e.g.
one can find misleading claims on the limited validity of the FDR:
we can cite an important review on turbulence containg 
 the following (wrong) conclusion 
{\it This absence of correlation between fluctuations and  relaxation
is reflected in the non existence of a fluctuation-dissipation theorem for
turbulence}~\cite{RS78}.

On the contrary, as we will discuss in the following, a generalized
FDR holds under rather general hypotheses~\cite{DH75,FIV90,BLMV03}.

\subsection{Chaos and the FDR: van Kampen's objection}

It has been argued by van Kampen~\cite{vK71} that the usual derivation of the FDR
that relies on a first-order truncation of the time-dependent
perturbation theory, for the evolution of probability density, can
been severely criticized.  In a nutshell, using the dynamical systems
terminology, the van Kampen's argument is as follows.  Given a
perturbation $\delta {\bf x}(0)$ on the state of the system at time
$t=0$, one can write a Taylor expansion for $\delta {\bf x}(t)$, the
difference between the perturbed trajectory and the unperturbed one:
\begin{equation}
\label{3.1}
\delta  x_i(t)=
\sum_{j} \frac{\partial x_i(t)}{\partial x_j(0)} \delta  x_j(0)
+O(|\delta {\bf x}(0)|^2) \,\,\, .
\end{equation}
Averaging over the initial condition one has the mean response function:
\begin{equation}
\label{3.2}
R_{i,j}(t)= \Biggl \langle \frac{\delta x_i(t)} {\delta x_j(0)}
\Biggr \rangle =\int \frac{\partial x_i(t)} {\partial x_j(0)}
\rho({\bf x}(0)) d{\bf x}(0 ) \,\,.
\end{equation}

In the case of the equilibrium statistical mechanics we have
$\rho({\bf x}) \propto \exp\Bigl( -\beta H(\bf{x}) \Bigr)$ so that ,
after an integration by parts, one obtains
\begin{equation}
\label{3.3}
R_{i,j}(t)= \beta \Biggl \langle x_i(t)
 \frac {\partial  H({\bf x}(0))} 
{ \partial x_j(0)} \Biggr \rangle, 
\end{equation}
which is nothing but the differential form of the usual FDR.

 In the presence of chaos the terms ${\partial x_i(t)}/{\partial
   x_j(0)}$ grow exponentially as $e^{\lambda t}$, where $\lambda$ is
 the Lyapunov exponent, therefore it is not possible to use the linear
 expansion (\ref{3.1}) for a time larger than $(1/ \lambda) \ln
 (L/|\delta{\bf x}(0)|)$, where $L$ is the typical fluctuation of the
 variable ${\bf x}$.  On account of that estimate, the linear response
 theory is expected to be valid only for extremely small and
 unphysical perturbations (or times).  For instance, according to the
 argument by van Kampen, if one wants that the FDR holds up to $1 s$
 when applied to the electrons in a conductor then a perturbing
 electric field should be smaller than $10^{-20} V/m$, in clear
 disagreement with the experience.

This result, at first glance,  sounds as bad news for the
statistical mechanics. 
However this criticism had the merit to stimulate 
for a deeper understanding of the basic ingreedients for the
validity of the FDR relation  and its validity range. 
On the other hand the success of the linear theory, for the computation
of the transport coefficients (e.g. electric conductivity) in terms of
correlation function of the unperturbed systems, is transparent, and its
validity has been, directly and undirectly, tested in a huge number of cases.
Therefore it is really
difficult to believe that the FDR cannot be applied for physically relevant
 value of the perturbations.
 
Kubo suggested that the origin of this effectiveness of the
Linear-Response theory may reside in the ``constructive role of
chaos'' because, as Kubo suggested, ``{\it instability [of the
    trajectories] instead favors the stability of distribution
  functions, working as the cause of the mixing}''~\cite{K86}.
The following derivation~\cite{FIV90} of a generalized FDR supports this intuition.

\subsection{Generalized FDR for stationary systems}
\label{Generalized FDR: stationary systems}

Let us briefly discuss a derivation, under rather general hypothesis,
of a generalized FDR (also in non equilibrium or non Hamiltonian
systems), for which the van Kampen critique does not hold.  Consider a
dynamical system $ {\bf x}(0) \to {\bf x}(t)=U^t {\bf x}(0)$ with
states ${\bf x}$ belonging to a $N$-dimensional vector space.  For the
sake of generality, we will consider the case in which the time
evolution can also be not completely deterministic (\textit{e.g.}
stochastic differential equations).  We assume the existence of an
invariant probability distribution $\rho({\bf x})$, for which some
``absolutely continuity'' type conditions are required (see later),
and the mixing character of the system (from which its ergodicity
follows).  Note that no assumption is made on $N$.

Our aim is to express the average response of a generic observable $A$
to a perturbation, in terms of suitable correlation functions,
computed according to the invariant measure of the unperturbed system.
At the first step we study the behaviour of one component of ${\bf  x}$,
 say $x_i$, when the system, described by $\rho({\bf x})$, is
subjected to an initial (non-random) perturbation such that 
${\bf  x}(0) \to {\bf x}(0) + \Delta {\bf x}_{0}$ \footnote{The study of an
  ``impulsive'' perturbation is not a  limitation, e.g. in
  the linear regime from the (differential) linear response one
  understands the effect of a generic perturbation.}.  This
instantaneous kick modifies the density of the system into
$\rho'({\bf x})$, related to the invariant
distribution by $\rho' ({\bf x}) = \rho ({\bf x} - \Delta {\bf x}_0)$.
We introduce the probability of transition from ${\bf x}_0$ at time
$0$ to ${\bf x}$ at time $t$, $w ({\bf x}_0,0 \to {\bf x},t)$. For a
deterministic system, with evolution law $ {\bf x}(t)=U^{t}{\bf
  x}(0)$, the probability of transition reduces to $w ({\bf x}_0,0 \to
{\bf x},t)=\delta({\bf x}-U^{t}{\bf x}_{0})$, where $\delta(\cdot)$ is the
Dirac's delta.  Then we can write an expression for the mean value of
the variable $x_i$, computed with the density of the perturbed system:
\begin{equation}
\label{3.4}
\Bigl \langle x_i(t) \Bigr \rangle ' = 
\int\!\int x_i \rho' ({\bf x}_0) 
w ({\bf x}_0,0 \to {\bf x},t) \, d{\bf x} \, d{\bf x}_0  \; .
\end{equation}
The mean value of $x_i$ during the unperturbed evolution can be written in
a similar way:
\begin{equation}
\label{3.5}
\Bigl \langle x_i(t) \Bigr \rangle = 
\int\!\int x_i \rho ({\bf x}_0) 
w ({\bf x}_0,0 \to {\bf x},t) \, d{\bf x} \, d{\bf x}_0  \; .
\end{equation}
Therefore, defining $\overline{\delta x_i} =  \langle x_i \rangle' -
\langle x_i \rangle$, we have:
\begin{eqnarray}
\label{3.6}
\overline{\delta x_i} \, (t)  &=&
\int  \int x_i \;
\frac{\rho ({\bf x}_0 - \Delta {\bf x}_0) - \rho ({\bf x}_0)}
{\rho ({\bf x}_0) } \;
\rho ({\bf x}_0) w ({\bf x_0},0 \to {\bf x},t) 
\, d{\bf x} \, d{\bf x}_0 \nonumber \\
&=& \Bigl \langle x_i(t) \;  F({\bf x}_0,\Delta {\bf x}_0) \Bigr \rangle
\end{eqnarray}
where
\begin{equation}
\label{3.7}
F({\bf x}_0,\Delta {\bf x}_0) =
\left[ \frac{\rho ({\bf x}_0 - \Delta {\bf x}_0) - \rho ({\bf x}_0)}
{\rho ({\bf x}_0)} \right] \; .
\end{equation}
Let us note here that the mixing property of the system is required so
that the decay to zero of the time-correlation functions assures the
switching off of the deviations from equilibrium.
 
For an infinitesimal perturbation $\delta {\bf x}(0) = (\delta x_1(0)
\cdots \delta x_N(0))$, if $\rho({\bf x})$ is non-vanishing and
differentiable, the function in (\ref{3.7}) can be expanded to first
order and one obtains:
\begin{eqnarray}
\label{3.8}
\overline{\delta x_i} \, (t)  &=&
- \sum_j \Biggl
\langle x_i(t) \left. \frac{\partial \ln \rho({\bf x})}{\partial x_j} 
\right|_{t=0}  \Biggr \rangle \delta x_j(0) \nonumber \\
&\equiv&
\sum_j R_{i,j}(t) \delta x_j(0)
\end{eqnarray}
which defines the linear response 
\begin{equation}
\label{3.9}
R_{i,j}(t) = - \Biggl \langle x_i(t) \left.
 \frac{\partial \ln \rho({\bf x})} {\partial x_j} \right|_{t=0}
\Biggr  \rangle 
\end{equation} 
of the variable $x_i$ with respect to a perturbation of $x_j$.

We note that in the above derivation  of the FDR relation we never
used any
approximation on the evolution of $\delta {\bf x}(t)$.  Starting with
the exact  expression (\ref{3.5}) for the response,
only a linearization on the initial time perturbed density is needed,
and this implies nothing but the smallness of the initial
perturbation.
We have to stress again that, from the evolution
of the trajectories difference, one can define the leading Lyapunov
exponent $\lambda$ by considering the absolute values of $\delta {\bf x}(t)$: 
at small $|\delta {\bf x}(0)|$ and large enough $t$ one has
\begin{equation}
\label{3.11}
\Bigl \langle \ln |\delta {\bf x}(t)|\Bigr \rangle \simeq 
\ln |\delta {\bf x}(0)| + \lambda t \,\, .
\end{equation}
On the other hand, in the FDR issue one deals with averages of
quantities with sign, such as $\overline{\delta {\bf x}(t)}$. 
This apparently marginal difference is very important and it is at the
basis of the possibility to derive the FDR relation avoiding the
van Kampen's objection.  

\subsection{Remarks on the invariant measure} 
\label{Stationary systems: remarks on the invariant measure} 
At this point one could object that in a chaotic deterministic
dissipative system the above machinery cannot be applied, because the
invariant measure is not smooth at all\footnote{Typically the invariant measure of a chaotic attractor has a
  multifractal character and its Renyi dimensions $d_q$ are not
  constant.}.  In chaotic dissipative systems the invariant measure is
singular, however the previous derivation of the FDR relation is still
valid if one considers perturbations along the expanding directions.
A general response function has two contributions, corresponding
respectively to the expanding (unstable) and the contracting (stable)
directions of the dynamics. The first contribution can be associated
to some correlation function of the dynamics on the attractor
(i.e. the unperturbed system). On the contrary this is not true for
the second contribution (from the contracting directions), this part
to the response is very difficult to extract numerically~\cite{CS07}.
Nevertheless a small amount of noise, that is always present in a
physical system, smoothes the $\rho({\bf x})$ and the FDR relation can
be derived.  We recall that this ``beneficial'' noise has the
important role of selecting the natural measure, and, in the numerical
experiments, it is provided by the round-off errors of the
computer. We stress that the assumption on the smoothness of the
invariant measure allows one to avoid subtle technical difficulties.

In Hamiltonian systems, taking the canonical ensemble as the
equilibrium distribution, one has
$ \ln \rho= -\beta H({\bf Q},{\bf P})\, + constant.$
Recalling Eq.~\eqref{3.9}, if we indicate there by $x_i$ the component $q_k$ of the
position vector ${\bf Q}$ and by $x_j$ the corresponding component $p_k$
of the momentum ${\bf P}$, from the Hamilton's equations
($dq_k/dt=\partial H/\partial p_k$) one has
the differential form of the usual FDR~\cite{K66,K86}
\begin{equation}
\label{3.12}
\frac{ \overline{\delta q_k} \, (t)} {\delta p_k(0)}
=\beta \Biggl \langle q_k(t) \frac {dq_k(0)} {dt} \Biggr \rangle =
- \beta \frac{d}{dt} \Bigl \langle q_k(t) q_k(0) \Bigr \rangle \, .
\end{equation}

When the stationary state is described by a multivariate Gaussian distribution, 
\begin{equation}
\ln \rho({\bf x})= -{1 \over 2} \sum_{i,j}\alpha_{ij}x_i x_j + const.
\end{equation}
with $\{ \alpha_{ij} \}$ a positive matrix, and 
the elements of the linear response matrix can be written in terms
of the usual  correlation functions:
\begin{equation}
\label{3.14}
R_{i,j} (t) =\sum_k \alpha_{i,k}
{\Bigl \langle x_i(t) x_k(0) \Bigr \rangle }  \; .
\end{equation}
The above result is nothing but the Onsager regression originally
obtained for linear Langevin equations.
It is interesting to note that there are 
important nonlinear  systems with a Gaussian invariant
distribution, e.g. the inviscid hydrodynamics~\cite{Kr59,Kr00}, where the Liouville theorem
holds, and a quadratic invariant exists.

In non Hamiltonian systems, where usually the shape of $\rho({\bf x})$
is not known, relation (\ref{3.9}) does not give a very detailed
quantitative information. 
However from it one can see  the existence of a connection between
the mean response function $R_{i,j}$ and a suitable correlation
function, computed in the non perturbed systems:
\begin{equation}
\label{3.13}
\Bigl \langle x_i(t)f_j({\bf x}(0)) \Bigr \rangle \; , 
\quad \textrm{with} \quad 
f_j({\bf x})=- \frac{\partial \ln \rho}{\partial x_j} \,\, ,
\end{equation}
where, in the general case, the function $f_j$ is unknown.  

Let us stress that in spite of the technical difficulty for the
determination of the function $f_j$, which depends on the invariant
measure, a FDR always holds in mixing systems whose invariant measure
is ``smooth enough''.  We note that the nature of the statistical
steady state (either equilibrium, or non equilibrium) has no role at
all for the validity of the FDR.

We close this Section noting that, as well clear even in the case
of Gaussian variables, the knoweledge of a marginal distribution
\begin{equation}
p_i(x_i)= \int \rho(x_1, x_2, ....)\prod_{j \neq i}dx_j
\label{projection} 
\end{equation} 
 is not enough for the computation of the auto response:
\begin{equation}
R_{i,i}(t)\neq
- \Biggl \langle x_i(t) \left.
 \frac{\partial \ln p_i(x_i)} {\partial x_i} \right|_{t=0}
\Biggr  \rangle \,\, , \label{falseresponse}
\end{equation}
This observation is particularly relevant for what follows.  Consider
that the description of our system has been restricted to a single
\emph{slow} degree of freedom, for instance the coordinate of a
colloidal particle ruled by the Langevin equation.  The above
discussion has indeed made clear that there can be other fluctuating
variables coupled to the one we are interested in, and it is not
correct to project them out by the marginalization in
Eq.~(\ref{projection}).  Conversely, a stationary propability
distribution with new variables coupled to the colloidal particle
position must be taken into account. In the examples discussed in the
last two sections,~\ref{Langevin process} and~\ref{GranularIntruder}
these additional fluctuating variables turn out to be local fields
coupled with the probe particle.

\subsection{Generalized FDR for Markovian systems}
\label{Generalized FDR: Markovian systems}

In the previous section we have discussed a formula useful to write an
explicit expression for the response to an external perturbation at
\emph{stationarity} in a case where the invariant measure is
known. The possibility to work out explicitly a response formula is
indeed quite general. In particular, for Markov processes a general
FDR has been derived~\cite{LCZ05,BMW09,CLSZ10}, and also
experimentally verified~\cite{GPCCG09,GPCM11}, which also holds for
non-stationary, aging processes, even in the absence of detailed
balance. Here we report a derivation which strictly follows the one
given in~\cite{LCZ05}.  An interesting outcome for the purpose of the
whole discussion will be the evidence that the response function
includes quite generally the correlation of the field of interest with
a local field.


Let us briefly recall that a Markov process is univocally defined by
the initial distribution probability $\rho({\bf x},0)$ and the
transition rates $W({\bf x'}\to{\bf x})$ from state ${\bf x'}$ to
state ${\bf x}$, with normalization

\begin{equation}
\sum_{{\bf x}} W({\bf x'}\to {\bf x})=0.
\label{2.0}
\end{equation}
The two-time conditional probability is
then approximated by
\begin{equation}
P({\bf x},t+\Delta t|{\bf x'},t)=\delta_{{\bf x},{\bf x'}}+
W({\bf x'}\to {\bf x})\Delta t + \mathcal{O}(\Delta t^2),
\label{2.1}
\end{equation}
where $\Delta t$ is a small time interval.  The average of a given
observable $A({\bf x})$ at time $t$ reads

\begin{equation}
\langle A(t)\rangle=\sum_{{\bf x},{\bf x'}}A({\bf x}) P({\bf x},t|{\bf
  x'},t')\rho({\bf x'},t').
\label{2.2}
\end{equation}


Now we discuss a perturbation to our system in the form of a
time-dependent external field $h(s)$, which couples to the potential
$V({\bf x})$ and changes the energy of the system from
$\mathcal{H}({\bf x})$ to $\mathcal{H}({\bf x})-h(s)V({\bf x})$. The
dynamics in the presence of the pertubation is decribed by a new
Markov process defined through the perturbed transition rates
$W^h({\bf x}|{\bf x'})$. There is not a univocal prescription
constraining on the choice of a particular form of $W^h$. Here we
focus on the FDR following from the particular choice which obeys the
so-called \emph{local detailed balance}~\cite{LB57}. In this case,
to linear order in $h$, the perturbed transition rates read, for ${\bf
  x}\ne {\bf x'}$,

\begin{equation}
W^h({\bf x'}\to {\bf x})=W({\bf x'}\to {\bf x})
\left\{1-\frac{\beta h}{2}\left[V({\bf x})-V({\bf x'})\right]+M({\bf x},{\bf x'})\right\},
\label{2.3}
\end{equation}
where $\beta$ is the inverse temperature and $M({\bf x},{\bf x'})$ is
an arbitrary function of order $\beta h$, symmetric with respect the
exchange of its arguments. The diagonal elements are obtained imposing
the normalization condition~(\ref{2.0}). Notice that, once the local
detailed balance is imposed, there remains a further degree of
arbitrariness in the choice of the function $M$. The dependence of the
FDR on the form of such function is studied
in~\cite{LCZ05,CLSZ10}. Here we focus on the particular case $M=0$.


Let us consider an impulsive perturbation turned on at time $s$ for
the time interval $\Delta t$. The linear response function of the
observable $A$ is, at $t>s+\Delta t$, with $t-s \gg \Delta t$,
\begin{equation}
R(t,s)= \left .\lim_{\Delta t\to 0}\frac{1}{\Delta t}\frac{\delta
  \langle A(t)\rangle_h}{\delta h(s)}\right|_{h=0},
\label{2.4}
\end{equation}
where $\langle\ldots\rangle_h$ denotes an average on the perturbed
dynamics. The derivative with respect to $h$ is written in terms of
the conditional probabilities as follows

\begin{equation}
 \left .\frac{\delta
  \langle A(t)\rangle_h}{\delta h(s)}\right|_{h=0}
=\sum_{{\bf x},{\bf x'},{\bf x''}}A({\bf x})P({\bf x},t|{\bf x'},s+\Delta t)
\left .\frac{\delta P^h({\bf x'},s+\Delta t|{\bf x''},s)}{\delta h}\right|_{h=0}
P({\bf x''},s).
\label{2.5}
\end{equation}

In order to derive a FDR relating the response function to correlation
functions computed in the unperturbed dynamics, the derivative with
respect to the external field in the rhs of Eq.~(\ref{2.5}) has to be
worked out explicitly. To do this, we notice that, enforcing the
normalization condition~(\ref{2.0}), the transition rates can be
separated in diagonal and off-diagonal contributions, namely

\begin{equation}
W^h({\bf x'}\to {\bf x})=-\delta_{{\bf x},{\bf x'}}\sum_{{\bf x''}\ne {\bf x'}}W^h({\bf x'}\to {\bf x''})+
\left(1-\delta_{{\bf x},{\bf x'}}\right)W^h({\bf x'}\to {\bf x}).
\label{2.6}
\end{equation}
Then, using Eqs.~(\ref{2.1}), (\ref{2.3}) and~(\ref{2.6}), one obtains

\begin{eqnarray}
&&\left .\frac{1}{\Delta t}\frac{\delta P^h({\bf x},s+\Delta t|{\bf x'},s)}{\delta h}\right|_{h=0}=
\nonumber \\
&&\frac{\beta}{2}\left\{\delta_{{\bf x},{\bf x'}}\sum_{{\bf x''}\ne {\bf x'}}
W({\bf x''}|{\bf x'})[V({\bf x'})-V({\bf x''})]
+(1-\delta_{{\bf x},{\bf x'}})W({\bf x'}\to {\bf x})[V({\bf x})-V({\bf x'})]\right\}. \nonumber \\
\label{2.7}
\end{eqnarray}
Such separation allows us to single out two different terms.
Indeed, substituting Eq.~(\ref{2.7}) into Eq.~(\ref{2.5}), 
one obtains

\begin{eqnarray}
  \left .\frac{\delta \langle A(t)\rangle_h}{\delta
   h(s)}\right|_{h=0} &=&\frac{\beta}{2}\sum_{{\bf x},{\bf x'},{\bf x''}}A({\bf
   x})P({\bf x},t|{\bf x'},s+\Delta t) 
\Big\{\Delta t W({\bf x'}\to {\bf x''})[V({\bf x'})-V({\bf x''})]P({\bf x'},s) \nonumber \\
&+& \Delta t W({\bf x''}\to {\bf x'})[V({\bf x'})-V({\bf x''})]P({\bf x''},s)
\Big\}.
\label{2.8}
\end{eqnarray}
Then, exploiting the time translation invariance of the conditional
probability, namely $P({\bf x},t+\Delta t|{\bf x'},t)=P({\bf x},t|{\bf
  x'},t-\Delta t)$, in the first line of the above equation, and using
Eq.~(\ref{2.1}) in the second one, in the limit $\Delta t\to 0$ one
obtains the response function
\begin{equation}
R(t,s)= \frac{\beta}{2}\left[\frac{\partial \langle A(t)
    V(s)\rangle}{\partial s} -\langle A(t)B(s)\rangle \right],
\label{2.9}
\end{equation}
where
\begin{equation}
B(s) \equiv B[{\bf x}(s)]=\sum_{{\bf x''}}\{V({\bf x''})-V[{\bf x}(s)]\}W[{\bf x}(s)\to {\bf x''}]
\label{2.10}
\end{equation}
is an observable quantity, namely depends only on the state of the
system at a given time. The relation~(\ref{2.9}) is known since a long
time in the context of overdamped Langevin equation for continuous
variables~\cite{CKP94}.  

For instance, in the case of the
  overdamped Langevin dynamics of a colloidal particle diffusing in a
  space dependent potential $U(x)$,
\begin{equation} 
\dot{x}(t) = - \frac{\partial U(x)}{\partial x}+ \sqrt{2 T } \zeta(t), 
\label{2.10bis}
\end{equation}
with $\zeta(t)$ a white noise with zero mean and unit variance, one
can derive a response formula, with respect to a perturbing force $F$,
analogous to Eq.~(\ref{2.9}) which reads as:
\begin{equation}
\frac{\delta \langle x(t) \rangle_F}{\delta F(s)} = 
\frac{\beta}{2} \left[ \frac{\partial \langle x(t) x(s) \rangle}{\partial s} - \langle x(t) B[x(s)] \rangle \right], 
\label{2.12}
\end{equation}
with $B[x(s)] = -\partial U/\partial x|_{x(s)}$. 
At equilibrium it can be easily proved that 
$\langle x(t) B[x(s)] \rangle  = - \partial \langle x(t) x(s) \rangle / \partial s $, 
recovering the standard FDT formula. Differently, out of equilibrium 
the contribution coming from the local field $B[x(s)]$ must be explicitly taken into account, 
as will be shown in examples in the next sections.

\section{Random walk on a comb lattice}
\label{Comb model}

\subsection{Anomalous diffusion and FDR}

As discussed above for Brownian motion, in the absence of external forcing one has, for large times $t\to\infty$,
\begin{equation}
\label{1}
\langle x(t) \rangle=0 \,\,\,\,\, ,  \,\,\,\,\,
\langle x^2(t) \rangle \simeq 2 D t \,\, ,
\end{equation}
where $x$ is the position of the Brownian particle and
the average is taken over the unperturbed dynamic. Once a small
constant external force $F$ is applied one has a linear drift
\begin{equation}
\label{2}
\overline{\delta x}(t)=
\langle x(t) \rangle_F - \langle x(t) \rangle
 \simeq \mu F t \,\,
\end{equation}
where $\langle\ldots\rangle_F$ indicates the average on the
perturbed system, and $\mu$ is the mobility of the colloidal particle.
 It is  remarkable that $\langle x^2(t) \rangle$
is proportional to $\overline{\delta x}(t)$ at any time:
\begin{equation}
\label{3}
{\langle x^2(t) \rangle  \over \overline{\delta x}(t)}=\frac{2}{\beta F},
\end{equation}
and the Einstein relation (see Eq.~\ref{EinsteinRelation}) holds.

On the other hand it is now well established that beyond the standard
diffusion, as in~(\ref{1}), one can have systems with anomalous
diffusion (see for instance~\cite{BG90,GSGWS96,CMMV99,MK00,BC05,BO02}), i.e.
\begin{equation}
\label{4}
\langle x^2(t)\rangle \sim t^{2 \nu} \,\,\, \mbox{with} \,\,\,  \nu \ne
1/2.
\end{equation}
Formally this corresponds to have $D=\infty$ if $\nu > 1/2$
(superdiffusion) and $D=0$ if $\nu < 1/2$ (subdiffusion).  In the
following we will limit the study to the case $\nu < 1/2$.  

It is quite natural to wonder if (and how) the FDR changes in the
presence of anomalous diffusion, i.e. if instead of~(\ref{1}),
Eq.~(\ref{4}) holds.  In some systems it has been shown
that~(\ref{3}) holds even in the subdiffusive case.  This has been
explicitly proved in systems described by a fractional-Fokker-Planck
equation~\cite{MBK99}, see also~\cite{BF98,CK09}. In addition there is
clear analytical~\cite{LATBL10} and numerical~\cite{VPV08} evidences
that~(\ref{3}) is valid for the elastic single file, i.e. a gas of
hard rods on a ring with elastic collisions, driven by an external
thermostat, which exhibits subdiffusive behavior, $\langle x^2 \rangle
\sim t^{1/2}$~\cite{HKK96}.

Here we discuss the validity of the fluctuation-dissipation relation
in the form~(\ref{3}) for a system with anomalous diffusion which is
not fully described by a fractional Fokker-Planck equation.  In
particular we will investigate the relevance of the anomalous
diffusion, the presence of non equilibrium conditions and the
(possible) role of finite size.  The example discussed here is the
diffusion of a particle on a comb lattice.  The dynamics of the model
is defined by transition rates and therefore a straightforward
application of the generalized FDR introduced in
Sections~\ref{Generalized FDR: Markovian systems}, Eq~(\ref{2.9}), is possible.

\subsection{The transition rates of the model}
\label{Model definition: transition rates}

The comb lattice is a discrete structure consisting of an infinite
linear chain (backbone), the sites of which are connected with other
linear chains (teeth) of length $L$~\cite{R01}. We denote by
$x\in(-\infty,\infty)$ the position of the particle performing the
random walk along the backbone and with $y\in[-L,L]$ that along a
tooth. The transition probabilities from state ${\bf x}\equiv(x,y)$ to
${\bf x'}\equiv(x',y')$ are:
\begin{eqnarray}
W^d[(x,0)\rightarrow (x\pm 1,0)]&=&1/4\pm d \nonumber \\
W^d[(x,0)\rightarrow (x,\pm 1)]&=&1/4 \nonumber \\
W^d[(x,y)\rightarrow (x,y\pm 1)]&=&1/2~~~ \textrm{for}~y\ne 0,\pm L.
\label{ww}
\end{eqnarray}
On the boundaries of each tooth, $y=\pm L$, the particle is reflected
with probability 1. The case $L=\infty$ is obtained in
numerical simulations by letting the $y$ coordinate increase without
boundaries. Here we consider a discrete time process and, of course,
the normalization $\sum_{(x',y')}W^d[(x,y)\rightarrow (x',y')]=1$
holds.  The parameter $d\in[0,1/4]$ allows us to consider also the
case where a constant external field is applied along the $x$ axis,
producing a non zero drift of the particle. A state with a non zero
drift can be considered as a perturbed state (in that case we denote
the perturbing field by $\varepsilon$), or it can be itself the
stationary state where a further perturbation can be added changing
$d\rightarrow d+\varepsilon$.

\subsection{Anomalous dynamics}
\label{Anomalous dynamics}
\begin{figure}
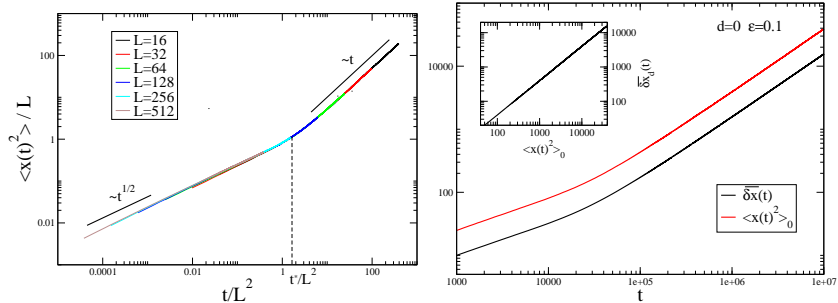

\includegraphics[width=.45\textwidth,clip=true]{comb1.eps}
\includegraphics[width=.45\textwidth,clip=true]{comb2.eps}
\caption{Left panel: $\langle x^2(t) \rangle_0/L$ \emph{vs}
$t/L^2$ is plotted for several values of $L$ in the comb model. Right
panel: $\langle x^2(t) \rangle_0$ and the response function
$\overline{\delta x}(t)$ for $L=512$. In the inset the
parametric plot $\overline{\delta x}(t)$ \emph{vs} $\langle x^2(t)
\rangle_0$ is shown.}
\label{fig1}
\end{figure}

Let us start by considering as a reference state the case $d=0$.  For
finite teeth length $L<\infty$, we have numerical evidence of a
dynamical crossover, at a time $t^*$, from a subdiffusive to a simple
diffusive asymptotic behaviour (see Fig.~\ref{fig1})
\begin{equation} 
\langle x^2(t) \rangle_0 \simeq
\left\{\begin{array}{cc}
C t^{1/2} &  t < t^{*}(L)\nonumber
\\ 
2 D(L) t &  t >t^{*}(L),
\label{nodrift2}
\end{array}\right.
\end{equation} 
where $C$ is a constant and $D(L)$ is an effective diffusion
coefficient depending on $L$. The symbol $\langle\ldots\rangle_0$
denotes an average over different realizations of the
dynamics~(\ref{ww}) with $d=0$ and initial condition $x(0)=y(0)=0$.
We find $t^*(L)\sim L^2$ and $D(L)\sim 1/L$. In the left panel of
Fig.~\ref{fig1} we plot $\langle x^2(t) \rangle_0/L$ as function of
$t/L^2$ for several values of $L$, showing an excellent data collapse.

In the limit of infinite teeth, $L \rightarrow \infty$,
$D\rightarrow 0$ and $t^*\rightarrow\infty$ and the system shows
a pure subdiffusive behaviour~\cite{HBA97} 
\begin{equation}
\langle x^2(t) \rangle_0
\sim t^{1/2}.
\label{nodrift1}
\end{equation} 
In this case, the probability distribution function behaves as
\begin{equation}
P_0(x,t)\sim t^{-1/4}e^{-c\left(\frac{|x|}{t^{1/4}}\right)^{4/3}},
\label{4/3}
\end{equation}
where $c$ is a constant, in agreement with an argument \emph{\`a la}
Flory~\cite{BG90}. The behaviour~(\ref{4/3}) also holds in the case of
finite $L$, provided that $t<t^*$. For larger times a simple Gaussian
distribution is observed. 


In the comb model with infinite teeth, the FDR in its standard form is
fulfilled, namely if we apply a constant perturbation $\varepsilon$
pulling the particles along the 1-d lattice one has numerical evidence that
\begin{equation} 
\langle x^2(t)\rangle_0 \simeq C\overline{\delta x}(t)\sim t^{1/2},
\label{FDRst}
\end{equation} 
where $\overline{\delta x}$ is the average in the presence of the
perturbation $\varepsilon$. Moreover, the proportionality between
$\langle x^2(t)\rangle_0$ and $\overline{\delta x}(t)$ is fulfilled
also with $L<\infty$, where both the mean square displacement (MSD)
and the drift with an applied force exhibit the same crossover from
subdiffusive, $\sim t^{1/2}$, to diffusive behavior, $\sim t$ (see
Fig.~\ref{fig1}, right panel).  Therefore what we can say is that the
FDR is somehow ``blind" to the dynamical crossover experienced by the
system.  When the perturbation is applied to a state without any
current, the proportionality between response and correlation holds
despite anomalous transport phenomena.

\subsection{Application of the generalized FDR}
\label{Application of the generalized FDR}

Our aim here is to show that, differently from what depicted above
about the zero current situation, within a state with a non zero drift
the emergence of a dynamical crossover is connected to the breaking of
the Einstein FDR~(\ref{EinsteinRelation}).  Indeed, the MSD in the
presence of a non zero current, even with $L=\infty$, shows a
dynamical crossover
\begin{align}
\langle x^2(t) \rangle_d &\sim  a~t^{1/2} + b~t, \label{drift1} \\
\langle [x(t)-\langle x(t)\rangle_d]^2 \rangle_d &\sim a'~t^{1/2}+b'~t,
\label{deltax2comb}
\end{align}
where $a$, $b$, $a'$ and $b'$ are constants, whereas
\begin{equation} 
\overline{\delta x}_d(t) \sim t^{1/2},
\label{drift2}
\end{equation}
with $\overline{\delta x}_d(t)=\langle
x(t)\rangle_{d+\varepsilon}-\langle x(t)\rangle_d$: hence at large times
the Einstein relation breaks down (see Fig.~\ref{figFDR}).  Notice that the
proportionality between response and fluctuations cannot be recovered
by simply replacing $\langle x^2(t) \rangle_d$ with $\langle x^2(t)
\rangle_d-\langle x(t)\rangle^2_d$, as it happens for Gaussian
processes (see discussion below).  The constants $a'$ and $b'$ can be computed
analitically in the case $L=\infty$~\cite{barkai}.


The first moment $\langle
x(t)\rangle_{d+\varepsilon}$ of $P_{d+\varepsilon}(x,t)$ is not proportional to
the second cumulant of $P_{d}(x,t)$, namely $ \langle x^2(t)\rangle_d-\langle
x(t)\rangle^2_d$.  In order to point out a relation between such
quantities, we need a generalized fluctuation-dissipation relation.

According to the definition~(\ref{ww}), one has for the backbone
\begin{eqnarray}
W^{d+\varepsilon}[(x,y)\rightarrow (x',y')] &=& W^d[(x,y)\rightarrow (x',y')]
\left(1+\frac{\varepsilon (x'-x)}{W^0+d (x'-x)}\right)\\
&\simeq& W^de^{\frac{\varepsilon}{W^0}(x'-x)}, \nonumber \\
\end{eqnarray}
where $W^0=1/4$, and the last expression holds under the condition
$d/W^0\ll 1$.  The above expression can be rewritten in the form of a
\emph{local detailed balance}, Eq.~(\ref{2.3}), with $V({\bf x})=x$
and $\beta=1$, yielding, for $(x,y)\ne (x',y')$,
\begin{equation}
W^{d+\varepsilon}[(x,y)\rightarrow (x',y')]=W^d[(x,y)\rightarrow (x',y')] 
e^{\frac{h(\varepsilon)}{2}(x'-x)},
\label{Wpert}
\end{equation}
where $h(\varepsilon)=2\varepsilon/W^0$. Using Eqs.~(\ref{2.9}) and~(\ref{2.10})
we obtain 
\begin{equation}
\frac{\overline{\delta\mathcal{O}}_d}{h(\varepsilon)}
=\frac{\langle \mathcal{O}(t)\rangle_{d+\varepsilon}- \langle
\mathcal{O}(t)\rangle_d}{h(\varepsilon)}
=\frac{1}{2}\left[\langle\mathcal{O}(t)x(t)\rangle_d-\langle\mathcal{O}(t)x(0)\rangle_d
-\langle\mathcal{O}(t)A(t,0)\rangle_d\right],
\label{FDR}
\end{equation}
where $\mathcal{O}$ is a generic observable, and $A(t,0)=\sum_{t'=0}^t
B(t')$, with, in this case,
\begin{equation}
B[(x,y)]=\sum_{(x',y')}(x'-x)W^d[(x,y)\rightarrow (x',y')].
\end{equation}
Recalling the definitions~(\ref{ww}), from the above equation we have
$B[(x,y)]=2d\delta_{y,0}$ and therefore the sum on $B$ has an
intuitive meaning: it counts the time spent by the particle on the $x$
axis.  The results described in section~\ref{Anomalous dynamics} can be then
read in the light of the fluctuation-dissipation relation~(\ref{FDR}):

\begin{figure}[t!]
\centering
\includegraphics[scale=0.4,clip=true]{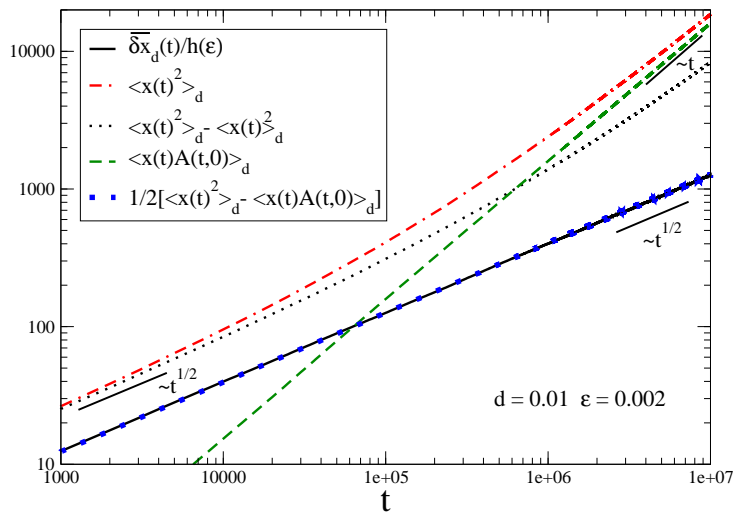}
\caption{Response function (black line), MSD (red dotted line) and
  second cumulant (black dotted line) measured in the the comb model
  with $L=\infty$, field $d=0.01$ and perturbation
  $\varepsilon=0.002$. The correlation involving the quantity $B$
  (green dotted line) yields the right correction to recover the full
  response function (blue dotted line), in agreement with the
  FDR~(\ref{FDR}).}
\label{figFDR}
\end{figure}

i) Putting $\mathcal{O}(t)=x(t)$, in the case without drift, i.e. $d=0$, one has $B=0$ and,
recalling the choice of the initial condition $x(0)=0$, 
\begin{equation}
\frac{\overline{\delta x}}{h(\varepsilon)}=\frac{\langle x(t)\rangle_{\varepsilon}- \langle x(t)\rangle_0}{h(\varepsilon)}=
\frac{1}{2}\langle x^2(t)\rangle_0.
\label{FDR1}
\end{equation}
This explains the observed behaviour (\ref{FDRst}) even in the
anomalous regime and predicts the correct proportionality factor,
$\overline{\delta x}(t)=\varepsilon/W^0\langle x^2(t)\rangle_0$.

ii) Putting $\mathcal{O}(t)=x(t)$, in the case with $d\ne 0$, one has
\begin{equation}
\frac{\overline{\delta x}_d}{h(\varepsilon)} =\frac{1}{2}\left[\langle
  x^2(t)\rangle_d-\langle x(t)A(t,0)\rangle_d\right].
\label{ale1}
\end{equation}
This explains the observed behaviours~(\ref{drift1})
and~(\ref{drift2}): the leading behavior at large times of $\langle
x^2(t)\rangle_d \sim t$, turns out to be exactly canceled by the term
$\langle x(t)A(t,0)\rangle_d$, so that the relation between response
and unperturbed correlation functions is recovered (see
Fig.~\ref{figFDR}).

iii) As discussed above, it is not enough to substitute $\langle
x^2(t)\rangle_d$ with $\langle x^2(t)\rangle_d-\langle
x(t)\rangle_d^2$ to recover the proportionality with $\overline{\delta
x}_d(t)$ when the process is not Gaussian. This can be explained in
the following manner. By making use of the second order out of equilibrium FDR derived
by Lippiello \emph{et al.} in~\cite{LCSZ08a,LCSZ08b,CLSZ10b}, which is needed due to the
vanishing of the first order term for symmetry, we can explicitly
evaluate
\begin{equation}
\langle x^2(t)\rangle_d
=\langle x^2(t)\rangle_0+h^2(d)\frac{1}{2}\left[\frac{1}{4}\langle x^4(t)\rangle_0
+\frac{1}{4}\langle x^2(t)A^{(2)}(t,0)\rangle_0\right],
\label{FDR2}
\end{equation}
where $A^{(2)}(t,0)=\sum_{t'=0}^t B^{(2)}(t')$ with $B^{(2)}=-\sum_{x'}(x'-x)^2W[(x,y)\rightarrow (x',y')]=-1/2\delta_{y,0}$.
Then, recalling Eq.~(\ref{FDR1}), we obtain
\begin{eqnarray}
&&\langle x^2(t)\rangle_d- \langle
x(t)\rangle^2_d= \nonumber \\
& & \langle
x^2(t)\rangle_0+h^2(d)\left[\frac{1}{8}\langle x^4(t)\rangle_0
+\frac{1}{8}\langle x^2(t)A^{(2)}(t,0)\rangle_0-\frac{1}{4}\langle
x^2(t)\rangle^2_0\right].
\label{FDR3}
\end{eqnarray}
Numerical simulations show that the term in the square brackets grows
like $t$ yielding a scaling behaviour with time consistent with
Eq.~(\ref{deltax2comb}).  On the other hand, in the case of the simple
random walk, one has $B^{(2)}=-1$ and $A^{(2)}(t,0)=-t$ and then
\begin{eqnarray}
&&\langle x^2(t)\rangle_d- \langle
x(t)\rangle^2_d= \nonumber \\
& & \langle x^2(t)\rangle_0+h^2(d)\left[\frac{1}{8}\langle x^4(t)\rangle_0
-\frac{1}{8}t\langle x^2(t)\rangle_0-\frac{1}{4}\langle
x^2(t)\rangle^2_0\right].
\label{FDR4}
\end{eqnarray}
Since in the Gaussian case $\langle x^4(t)\rangle_0=3\langle
x^2(t)\rangle^2_0$ and $\langle x^2(t)\rangle_0=t$, the term in the
square brackets vanishes identically and that explains why, in the
presence of a drift, the second cumulant grows \emph{exactly} as the
second moment with no drift.

\section{Entropy Production}
\label{Fluctuation Theorem}
  Non equilibrium regimes are always characterized by some sort of
  current flowing across the systems.  In the following sections we
  will suggest, showing examples, some connections between such
  non-equilibrium currents and the new degrees of freedom entering the
  fluctuation dissipation relation, while in the present one we
  introduce a general symmetry relation of the probability
  distribution of such currents, the Fluctuation relation.

In general, when one deals with non-equilibrium dynamics, very few
results independent from the details of the model are
available. Actually in the last decade a group of relations, known
with the name of ``fluctuation relations'' have captured the interest
of the scientific community, especially for the vast range of
applicability. Initially, a numerical evidence given by Evans and
collaborators showed a particular symmetry in the distribution of an
observable of a molecular fluid under shear. In a second moment, such
a symmetry has been proved as a theorem by Gallavotti and Cohen, under
quite general hypothesis.  The interested reader can see, among
others~\cite{review}. For a review of large deviation theory applied
to non-equilibrium systems, see~\cite{TH12}.

According to the  point of view here adopted we will focus on
systems in which some randomness is present.  Thanks to this
assumption, it is possible to skip several technical problems and some
forms of fluctuation theorems for stochastic systems can be
used. Among others, we are going to use the Lebowitz-Spohn functional,
valid for Markovian dynamics. In order to fix ideas, let us consider a
one-dimensional process discrete in time (the generalization to the
other cases is straightforward) and let us identify a trajectory in a
time interval $[0,t]$:
\begin{equation}
\Omega_{t}=\{x(0),x(1),\dots,x(t)\}
\end{equation}
Clearly, because of the stochastic nature of the dynamics it is
possible to associate to the trajectories a probability
$P(\Omega_{t})$; using the Markovian nature of the process one has
\begin{equation}
P(\Omega_{t})=p(x(0))\prod_{n=1}^{t}W(x(n-1)\to x(n))
\end{equation}

Analogously, one can consider the time-reversed trajectory, 
i.e. $\overline{\Omega}_{t}={x(t),x(t-1),\dots,x(0)}$ with
its probability $P(\overline{\Omega_{t}})$. For clarity, here we are considering variables which
are \emph{even} in time. This is not the most general case: for
example, when one deals also with velocities one must take into
account the different parity of the variables.
Then the Lebowitz-Spohn functional is defined as
\begin{equation}
\Sigma_{t}=\log{\frac{P(\Omega_{t})}{P(\overline{\Omega}_{t})}}. 
\label{entropyprod}
\end{equation}
Note that this quantity, in general, is not related to a specific
thermodynamic observable. However, in what follows, we will call it
``entropy production'' according to the recent literature. If the
dynamics satisfies detailed balance conditions, one has that
$P(\Omega_{t})=P(\overline{\Omega}_{t})$ and then $\Sigma_{t}$ is
identically equal to zero. On the contrary, when the detailed balance
condition is broken its average is strictly positive and does
increase in time
\begin{equation}
\langle \Sigma_{t} \rangle \sim t \phantom{mmmmmm}\textrm{for $t$ large}
\end{equation}
where the brackets mean an average on the space of trajectories in the stationary ensemble.  In
this sense, one of the main features of this quantity is that it
captures the ``non-equilibrium'' nature of the system.

Let us discuss with a pedagogical example how the entropy
  production is related to non-equilibrium currents. Consider a Markov
  process where the perturbation of an external force $F$ inducing
  non-equilibrium currents enters in the transition rates according to
  \emph{local detailed balance} condition
\begin{equation} 
\frac{W_F(x\rightarrow x')}{W_F(x'\rightarrow
  x)}=\frac{W_0(x\rightarrow x')}{W_0(x'\rightarrow x)} e^{2\beta F
   j(x\rightarrow x')},
\label{3.10}
\end{equation}  
with $j(x\rightarrow x')$ the current associated to the
transition $x\rightarrow x'$, which obey the symmetry property
$j(x\rightarrow x') = - j(x' \rightarrow x)$. According to the definition of
entropy production~(\ref{entropyprod}) one finds, for large times,
\begin{equation}
\frac{\Sigma_t}{t} = 2 \beta F \frac{1}{t} \sum_{n=1}^{t}J(x(n-1)\to
x(n)) = 2 \beta F J(t),
\end{equation}
where $J(t)$ is the average current over a time window of duration $t$. The fluctuation
theorem is a symmetry property of the probability distribution of the
variable $y=\Sigma_t/t$ which reads:
\begin{equation}
  \frac{P(y)}{P(-y)}=e^{y} \label{FT} \Longrightarrow \frac{P(2\beta
    FJ(t))}{P(-2\beta F J(t))}=e^{2\beta FJ(t)}.
\end{equation}
Namely the fluctuation theorem describes a symmetry in the
fluctuations of currents. Also, for large times we can assume a large
deviation hypothesis $P(y)\sim e^{-t S(y)}$, with $S(y)$ a Cramer
function. For small fluctuations around the mean value of $y$ the
Cramer function can be approximated to $S(y) = S(2 \beta F J) \simeq
\beta^2 F^2 (J-\overline{J})^2/\sigma_J^2$, where $\sigma_J$ is the
variance.  The fluctuation relation reads as $S(y)-S(-y)=y$; in the
Gaussian limit ($y$ close to $\overline{y}$) the previous constraint
can be easily demonstrated to be equivalent to $\overline{J}/F = \beta
\sigma_J^2$, which is nothing but the standard fluctuation dissipation
relation.  Therefore the fluctuation relation, which in some simple
situation can be directly related to the fluctuation-dissipation
relation, is a more general symmetry to which we expect to obey the
fluctuating entropy production. For a more general discussion of the
link between the Lebowitz-Spohn entropy production and currents,
see~\cite{AG07,LS99}.
The remarkable fact appearing in equation~(\ref{FT}) is that it does not
contain any free parameter, and so, in this sense, is model-independent.

\section{Langevin processes without detailed balance}
\label{Langevin process}

Sometimes the properties of statistical
  systems are investigated studying the diffusional properties of
  probe particles which customarily have larger mass and size compared
  with the constituents of the environment. This is for instance the
  case of Brownian motion, in which the erratic motion of flower-dust
  within water molecules was considered.  The central point is that
  such probe particle is always coupled to a \emph{small}
  portion of the whole system, so that small scale fluctuations always
  do matter.  We are going to study the feedback of local fluctuations
  of the fluid surrounding the probe on the dynamics of the probe
  itself when the environment is out of equilibrium.  The dynamics of
  a single probe particle in contact with an equilibrium thermal bath
  is well described by a single stochastic differential
  equation:
\begin{equation} 
M\dot{ V}(t)=-\int_{-\infty}^tdt'~\Gamma(t-t') V(t')+ {\cal E}(t'),
\label{langmemory}
\end{equation}
where ${\cal E}(t)$ is a gaussian white noise with zero mean and variance $\langle
{\cal E}(t){\cal E}(t')\rangle = \Gamma(|t-t'|)$ (second kind FDT).
The equilibrium condition is always guaranteed by the proportionality
between memory kernel and noise autocorrelation.  The simpler way to
account for the coupling between our probe particle with \emph{non
  equilibrium} fluctuations of the environments is by breaking the second
kind FDT.

In the rest of the section, we will discuss the following points:
\begin{itemize}
 \item how
non-equilibrium arises in a multidimensional linear Langevin model;
\item how the coupling between the variable of interest and others do matter
in the response formula;
\item what form is taken by the Fluctuation-Relation in such a system.
 \end{itemize}
\subsection{Markovian linear system}
\label{Markovian linear system}




The coupling between the probe and the portion of
fluid surrounding it, Eq.~\eqref{langmemory}, in certain regimes may be effectively described in terms of two
coupled Langevin equations: one variable is the velocity of our tagged
particle and the other is a local force field. Such a simplification can be realized in not too dense
cases, where the memory kernel $\Gamma$ has a single characteristic
finite time: a more detailed discussion can be found
in~\cite{VBPV09,PV09} and a specific example, in the context of granualr systems, is given in
section~\ref{GranularIntruder}.  The system can be then put in the
following form~\cite{SVGP10}:
\begin{eqnarray}
\label{eqmotion} M\dot{V} &=& -\Gamma(V-U)+\sqrt{2\Gamma T_1}\phi_1\\ \nonumber
M'\dot{U} &=& -\Gamma' U - \Gamma V+\sqrt{2 \Gamma' T_2} \phi_2,
\end{eqnarray}
where $M$ and $M'$ are ``masses'', $\Gamma$ and $\Gamma'$ are sort of
viscosities and $T_1$ and $T_2$ are two different ``energy scales''.
Model~\eqref{eqmotion}, in a more compact form, reads
\begin{equation}
\frac{d\bf X}{dt}=-A{\bf X}+ {\bf \phi}, \label{2main}
\end{equation}
where ${\bf X}\equiv(X_1,X_2)$ e ${\bf \phi}\equiv(\phi_1,\phi_2)$ are
2-dimensional vectors and $A$ is a real $2\times 2$ matrix
and ${\bf \phi}(t)$ a Gaussian process, with zero mean and covariance matrix:
\begin{equation}
\left<\phi_{i}(t')\phi_{j}(t)\right>=2\delta(t-t')D_{ij},
\end{equation}
and 
\begin{equation}
A=\left(
\begin{array}{cc}
\frac{\Gamma}{M} & -\frac{\Gamma}{M}  \\
 \frac{\Gamma}{M'}  & \frac{\Gamma'}{M'} 
\end{array}
\right)\phantom{mmmmm}
D=\left(
\begin{array}{cc}
  \frac{\Gamma T_1}{M^2} & 0  \\
0  & \frac{\Gamma' T_2}{(M')^2} 
\end{array}
\right) .
\end{equation}
The stability conditions on the dynamical matrix which guarantee
the existence of a stationary state are $\textrm{Tr}(A) > 0 $ and 
$\textrm{det}(A) > 0 $ which are clearly fulfilled in the present case. 
The invariant measure of the steady state is represented 
by a $2$-variate Gaussian distribution:
\begin{equation}
\rho({\bf X})=N\exp{\left(-\frac{1}{2}{\bf X}\sigma^{-1}{\bf X}\right)} \label{GaussNESS}
\end{equation}
where $N$ is a normalization coefficient and the matrix of covariances $\sigma$ 
is obtained by solving
\begin{equation}
D=A\sigma+\sigma A^{T},
\label{sigmamatrix}
\end{equation}
which yields:
\begin{equation}
\sigma=\left(
\begin{array}{cc}
\frac{T}{M}+\Theta \Delta T & \Theta \Delta T \\
\Theta \Delta T & \frac{T_{1}}{M'}+\frac{\Gamma}{\Gamma'}\Theta \Delta T \\
\end{array}
\right) \label{SigmaNESS}
\end{equation}
where
$\Theta=\frac{\Gamma\Gamma'}{(\Gamma+\Gamma')(M'\Gamma+M\Gamma')}$ and $\Delta T = T_{1}-T_{2}$. 
It is now well clear that when $T_1=T_2$ the two
variables are uncorrelated. This is the fingerprint of an equilibrium
condition, as we shall see in the following sections. 

\subsection{Fluctuation-response relation}
\label{Linear system: fluctuation-response relation}

We can now explicitly study the fluctuation and response properties of
the system since the dynamics is linear. First, the correlation matrix
$C_{ij} (t,t') =\langle X_{i} (t)X_{j} (t')\rangle$, in the stationary
state is time-translational invariant, i.e. it only depends on the
time difference $t-t'$. Then, using the equation of motion, it is
immediate to verify that $\dot{C}(t) = -AC(t)$, with initial condition
given by the covariance matrix, i.e. $C(0) = \sigma$. The
corresponding solution (in the matrix form) is: $C(t) = e^{-At}\sigma
$ Note that, in general $\sigma$ and $A$ do not commute. Moreover,
considering the gaussian shape of the steady state distribution
function~(\ref{GaussNESS}) it is possible to calculate the response
function of the system, by a straightforward application of
(\ref{3.14}):
\begin{equation}
{\bf R}(t)\equiv \overline{\frac{\delta X_i(t)}{\delta X_j(0)}}=
\sum_{i,j}\sigma^{-1}_{ij}\langle X_{i}(t)X_{j}(0)\rangle= {\bf
  C}(t)\sigma^{-1}. \label{GeneralizedResp}
\end{equation}
In the case of our interest, by considering a perturbation applied to
the variable $V$, one obtains:
\begin{equation}
\overline{\frac{\delta V(t)}{\delta V(0)}}=\sigma^{-1}_{11}\langle
V(t)V(0)\rangle+\sigma^{-1}_{21}\langle
V(t)U(0)\rangle. \label{GFRlangevin}
\end{equation}
As it appears clear from equation (\ref{GFRlangevin}), the response of
the variable $V$ to an external perturbation in general is not
proportional to the unperturbed autocorrelation.  It is possible to
observe the following scenario:  in the case $T_{1}=T_{2}\equiv
T$, i.e. $\Delta T= 0$, $\sigma$ is diagonal with
$\sigma_{11}=\frac{T}{M}$ and $\sigma_{22}=\frac{T}{M'}$,
independently of the values of other parameters, a direct
proportionality between $C_{VV}$ and $R_{VV}$ is obtained. This is not
the only case for this to happen: also for $\Gamma \ll \Gamma'$, or
$\Gamma\gg\Gamma'$, $\sigma_{12}$ goes to zero and a direct
proportionality between response and autocorrelation of velocity is
recovered.  More in general, when $T_{2}\neq T_{1}$, the coupling term
$\sigma_{12}$ differs from zero and a ``violation'' of the Einstein
relation emerges, triggered by the coupling between different degrees of freedom.

\begin{figure}
\begin{center}
\centering
\includegraphics[width=8cm,clip=true]{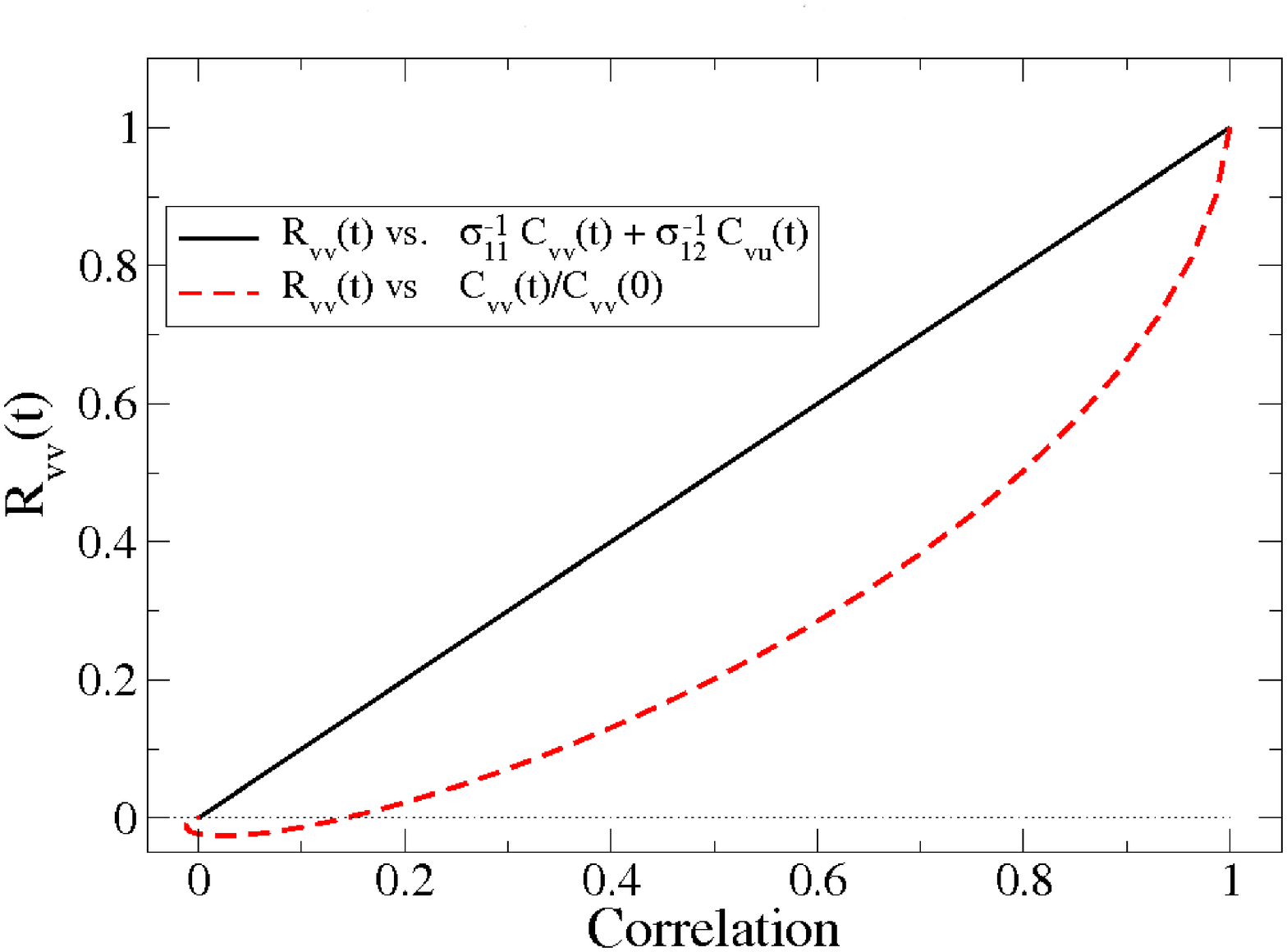}
\caption{Free particle with viscosity and memory, whose dynamics is
  given by Eq.~\eqref{eqmotion}: here we show the
  parametric plot of the velocity response to an impulsive
  perturbation at time $0$, versus two different correlations. The
  Einstein relation, which is not satisfied, would correspond to a
  linear shape with slope $1$ for the red dashed line. The black line
  shows that the Generalized response relation holds.
\label{fig:kubo}}
\end{center}
\end{figure}

The situation is clarified by Fig.~\ref{fig:kubo}, where the Einstein
Relation is violated and the generalized FDR holds: response $R_{VV}(t)$, when
plotted against $C_{VV}(t)$, shows a non-linear (and non-monotonic)
graph. Anyway a simple linear plot is restored when the response is
plotted against the linear combination of correlations indicated by
formula~\eqref{GFRlangevin}. In this case it is evident that the
``violation'' cannot be interpreted by means of any effective
temperature~\cite{CKP97}: on the contrary it is a consequence of having ``missed''
the coupling between variables $V$ and $U$, which gives an additive
contribution to the response of $V$.

The above consideration can be easily generalized to the case where
several auxiliary fields are present: let us suppose that there are
$N-1$ variables at temperature $T_{1}$ (the velocity $V$ and $N-2$
 auxiliary variables $U_{i}$) and one at temperature $T_{2}$. In this
case it is possible to show that the off-diagonal terms in $\sigma$
are proportional to $(T_{1}-T_{2})$ and a qualitative similar analysis
can be performed.

It is useful to stress the role of Markovianity, which is relevant for
a correct prediction of the response. In fact the marginal probability
distribution of velocity $P_{m}(V)$ can be computed straightforwardly
and has always a Gaussian shape. By that, one could be tempted to
conclude, inserting $P_{m}(V)$ inside , that proportionality between
response and correlation holds also if $T_{2}\neq T_{1}$. This
conclusion is wrong, as shown in equation (\ref{falseresponse})
because the process is Markovian only if both the variable $V$ and the
"hidden" variable $U$ are considered.

 We conclude by mentioning a sentence of Onsager: \emph{how do you
   know you have taken enough variables, for it to be Markovian?}~\cite{OM53}. We
 have seen in this section how a correct Response-Fluctuation analisys
 is intimately related to this important \emph{caveat}.



\subsection{Entropy production}
\label{entropy_production}

Due to the linearity of the model (\ref{2main}),
the explicit calculation of the entropy production according to 
the Lebowitz-Spohn definition is quite straightforward. 
There is a one to one mapping between the trajectories of the system 
and the realizations of the noise which allows to write the probability 
of each trajectory as a path-integral over all the possible realizations of noise:
\begin{equation}
 P(\{ {\bf X}(s)\}_0^t)=\int \mathcal{D}{\bf \phi} P({\bf
   \phi})\delta(\dot{{\bf X}}+A{\bf X}-\phi)\,.
 \end{equation}
Due to the gaussianity of $P(\phi)$ the above integral is easily evaluated
yielding the well know \emph{Onsager-Machlup} expression for the path probabilities:
\begin{equation}
P(\{ {\bf X}(s)\}_0^t)\sim e^{-(\dot{{\bf X}}+A{\bf
    X})D^{-1}(\dot{{\bf X}}+A{\bf X})}. 
\label{OnsagerMachlup}
 \end{equation}

In our model it is possible to compute the entropy production functional of a
single trajectory.  Let us consider a general trajectory $\{ {\bf
  X}(s)\}_0^t$ and its time-reversal $\{ \overline{{\bf
    X}}(s)\}_0^t$. Lebowitz and Spohn defined a fluctuating entropy production
functional $W_t$ as follows~\cite{LS99}:
\begin{equation}
\Sigma_t'=\log\frac{p[{\bf X}(0)]P(\{ {\bf X}(s)\}_0^t)}{p[{\bf X}(t)]P(\{
  \overline{{\bf X}}(s)\}_0^t)}=W_t+b_t 
\label{entropy_definition}
\end{equation}
with 
\begin{equation}
b_t=\log\{f[{\bf X}(0)]\}-\log\{f[({\bf X})]\},
\end{equation}
where $p[{\bf X}(0)]$ is the stationary distribution, i.e. a bivariate
gaussian with covariance given by
equation~(\ref{sigmamatrix}). Lebowitz and Sphon have shown that the
average (over the steady ensemble) of $W_{t}$, if detailed balance is
not satisfied, increases linearly with time, while the term $b_t$, usually
known as ``border term'', is usually negligible for large times,
unless particular conditions of ``singularity''
occur~\cite{othertheory,PRV06,BGGZ06}.

For semplicity of the notation, let us define
$\left(\begin{array}{c}F_{1}(\bX)\\F_{2}(\bX)\end{array}\right)\equiv
-A\bX$. In order to write down an explicit expression for the entropy
production, it is necessary to establish parity of variables under
time-reversal (i.e. positions are even and velocities are odd under
time inversion transformation). Let us assume that under time
reversal $\overline{X_{i}}=\epsilon_{i}X_{i}$, where $\epsilon_{i}$
can be $+1$ or $-1$, using also $\ebX \equiv (\epsilon_1 X_1,
\epsilon_2 X_2)$. Then it is possible to define
\begin{eqnarray} 
F_i^{rev}(\bX)&=&\frac{1}{2}[F_i(\bX)-\epsilon_i
  F_i(\ebX)]=-\epsilon_iF_i^{rev}(\ebX) \label{drev}\\ F_i^{ir}(\bX)&=&\frac{1}{2}[F_i(\bX)+\epsilon_i
  F_i(\ebX)]=\epsilon_iF_i^{ir}(\ebX) \label{dirr}.
\end{eqnarray}
Given this notation~\cite{R89} it is possible to write down a
compact form for the entropy production\footnote{Note that if the
  correlation matrix of the noise is not diagonal, this formula is
  slightly different.} by simply substituting equation
(\ref{OnsagerMachlup}) into (\ref{entropy_definition}) and obtaining:
\begin{equation} 
\Sigma_t = \sum_{k}D^{-1}_{kk}\int_{0}^{t}ds F^{ir}_{k}\left[
  \dot{X}_{k}-F^{rev}_{k}\right] \label{formulone2},
\end{equation}
where the sum is over $k$ such that $D_{kk}\ne 0$.  Formula
(\ref{formulone2}) is valid also in presence on non-linear terms and
with several variables~\cite{PV09}. 

For the model system in Eq.~(\ref{eqmotion}), it appears clear that
the variable $V$, being velocity, is odd under the change of time,
while variable $U$, which represents the local force field (divided by $\Gamma$), is
even.  According to the above formula the entropy production reads as:
\begin{equation}
\Sigma_t\simeq \Gamma\left(\frac{1}{T_{2}}-\frac{1}{T_{1}}\right)\int_{0}^{t}V(t')U(t')dt'. \label{EntropyTracer}
\end{equation}
Note that, from (\ref{EntropyTracer}), one recovers the equilibrium
case, i.e. $T_{1}=T_{2}$, for which there is no entropy
production. Remarkably, the equality of the two temperatures is the
same equilibrium condition given by the FDT analysis.

\section{Granular intruder}
\label{GranularIntruder}

Models of granular fluids~\cite{JNB96b} are an interesting framework
where the issues discussed in the previous sections can be
addressed. Due to dissipative interactions among the microscopic
constituents, energy is not conserved and external sources are
necessary in order to maintain a non-trivial stationary state: time reversal
invariance is broken and consequently, properties such as the
Equilibrium Fluctuation-Dissipation relation (EFDR) do not hold. In
recent years, a rather complete theory for fluctuations in granular
systems has been developed for the dilute limit, in good agreement
with numerical simulations~\cite{BP04,BMG09}. However, a general
understanding of dense cases is still lacking. A common approach is
the so-called Enskog correction~\cite{BP04,DS06}, which renormalizes
the collision frequency to take into account the breakdown of
Molecular Chaos due to high density. In cooling regimes, the Enskog
theory may describe strong non-equilibrium effects, due to the
explicit cooling time-dependence~\cite{SD01}. However it cannot
describe dynamical effects in stationary regimes, such as violations
of the Einstein relation~\cite{G04,PBV07}. Indeed, as discussed
before, violations of Einstein relation comes from having neglected
{\em coupled} degrees of freedom which are expected to be independent
at equilibrium. The Enskog approximation is not sufficient to describe
the effect of such a coupling, because it does not break factorization
of velocities:
\begin{equation}
\rho(x_1,...x_N,v_1,...v_N)=\rho_x(x_1,...,x_N)\prod_{i=1}^Np_1(v_i)
\end{equation}
with $p_1(v)$ the single-particle velocity distribution. Such an
approximation, which predicts the validity of Einstein relation, is
revealed to be wrong already at moderate densities in granular fluids,
as discussed in the following.

\subsection{Model}
\label{Model}

A good benchmark for new non-equilibrium theories is provided by the dynamics of a
massive tracer interacting with a gas of smaller granular particles, both
coupled to an external bath. In particular, taking as a reference point
the dilute limit, where the system has a closed analytical
description~\cite{SVCP10}, it is shown that more dense configurations
are well described by a Generalized Langevin Equation (GLE) with an
exponential memory kernel of the form given in Eq.~(\ref{langmemory}), at least as
a first approximation capable of describing violations of EFDR and
other non-equilibrium properties~\cite{SVGP10}. Here, the main
features are:
\begin{itemize}
 \item the decay of correlation and response functions is
not a simple exponential and shows backscattering~\cite{OK07,FAZ09};
\item the EFDR~\cite{KTH91,BPRV08} of the first and second kind do not
hold. 
\end{itemize}
 In the model described here detailed balance is not satisfied in
 general, non-equilibrium effects can be taken into account and the
 correct behavior of correlation and response functions is
 predicted. Furthermore, the model has a remarkable property: it can
 be mapped onto the two-variable Markovian system discussed in the
 previous section, i.e. two coupled Langevin equations, as in
 Eqs.~(\ref{eqmotion}).  The dilute limit is then naturally recovered
 by putting to zero the coupling constant between the original
 variable and the auxiliary one.  The auxiliary variable can be
 identified in the local velocity field spontaneously appearing in the
 surrounding fluid. This allows us to measure the fluctuating entropy
 production (see Eq.(\ref{entropyprod})~\cite{seifert05}, and fairly
 verify the Fluctuation Relation~(\ref{FT})
 ~\cite{Kurchan,LS99,BPRV08}, a remarkable result, if considered the
 interest of the community~\cite{BGGZ06b} and compared with
 unsuccessful past attempts~\cite{FM04,PVBTW05}.

The model considered here~\cite{SVGP10} is the following: an ``intruder'' disc of mass $m_0=M$ and radius $R$,
moving in a gas of $N$ granular discs with mass $m_i=m$ ($i>0$) and
radius $r$, in a two dimensional box of area $A=L^2$. We denote by
$n=N/A$ the number density of the gas and by $\phi$ the occupied
volume fraction, i.e. $\phi=\pi(Nr^2+R^2)/A$ and we denote by $\bm{V}$
(or $\bm{v}_0$) and $\bm{v}$ (or $\bm{v}_i$ with $i>0$) the velocity
vector of the tracer and of the gas particles, respectively.
Interactions among the particles are hard-core binary instantaneous
inelastic collisions, such that particle $i$, after a collision with
particle $j$, comes out with a velocity
\begin{equation}
\bm{v}_i'=\bm{v}_i-(1+\alpha)\frac{m_j}{m_i+m_j}[(\bm{v}_i-\bm{v}_j)\cdot\hat{\bm{n}}]\hat{\bm{n}}
\end{equation}
where $\hat{\bm{n}}$ is the unit vector joining the particles' centers
of mass and $\alpha \in [0,1]$ is the restitution coefficient
($\alpha=1$ is the elastic case).
The mean free path of the intruder is proportional to $l_0=1/((r+R)n)$ and we denote
by $\tau_c$ its mean collision time. Two kinetic temperatures can be
introduced for the two species: the gas granular temperature
$T_g=m\langle \bm{v}^2\rangle/2$ and the tracer temperature
$T_{tr}=M\langle \bm{V}^2\rangle/2$.

In order to maintain a granular medium in a fluidized state, an
external energy source is coupled to each particle in the form of a
thermal bath~\cite{WM96,NETP99,PLMPV98} (hereafter,
exploiting isotropy, we consider only one component of the
velocities):
\begin{equation}
m_i\dot{v}_i(t)=-\gamma_b v_i(t) + f_i(t) + \xi_b(t).
\label{langgas}
\end{equation}
Here $f_i(t)$ is the force taking into account the collisions of particle $i$ with other
particles, and $\xi_b(t)$ is a white noise (different for all particles), with
$\langle\xi_b(t)\rangle=0$ and $\langle\xi_{b}(t)\xi_{b}(t')\rangle
=2T_b\gamma_b\delta(t-t')$.  The effect of the external energy source
balances the energy lost in the collisions and a stationary state is
attained with $m_i\langle v_i^2 \rangle \leq T_b$ .

At low packing fractions, $\phi < 0.1$, and in the large mass limit,
$m/M\ll 1$, using the Enskog approximation it has been
shown~\cite{SVCP10} that the dynamics of the intruder is described by
a linear Langevin equation:
\begin{equation}
M\dot{V}=-\Gamma_E V+ {\cal E}_E,
\label{langtracer}
\end{equation}
where ${\cal E}_E$ is a white noise such that
\begin{equation}
 \langle
      {\cal E}_i(t){\cal E}_j(t')\rangle = 2\left[\gamma_b
        T_b+\gamma_g^{E}\left(
        \frac{1+\alpha}{2}T_g\right)\right]\delta_{ij}\delta(t-t')
\end{equation}
and
\begin{equation}
\Gamma_E=\gamma_b+\gamma_g^E, \quad \textrm{with} \quad
\gamma_g^E=\frac{g_2(r+R)}{l_0}\sqrt{2\pi mT_g}(1+\alpha)
\end{equation}
where $g_2(r+R)$ is the pair correlation function for a gas particle and
the intruder at contact. In this limit the velocity autocorrelation function shows a simple
exponential decay, with characteristic time $M/\Gamma_E$. Time-reversal and the EFDR, which are very
weakly modified for uniform dilute granular
gases~\cite{PBL02,G04,PVTW06}, become perfectly satisfied for a
massive intruder. The temperature of the tracer is computed as
$T_{tr}^E=(\gamma_bT_b+\gamma_g^E\frac{1+\alpha}{2}T_g)/\Gamma_E$. For
a general study of a Langevin equation with ``two temperatures'' but a
single time scale (which is always at equilibrium), see
also~\cite{V06}.

\subsection{Dense case: double Langevin with two temperatures}
\label{Dense case: double Langevin with two temperatures}

As the packing fraction is increased, the Enskog approximation is less
and less effective in predicting memory effects and the dynamical
properties.  In particular, velocity autocorrelation $C(t)=\langle
V(t)V(0)\rangle/\langle V^2\rangle$ and linear response function
$R(t)=\overline{\delta V(t)}/\delta V(0)$ (i.e. the mean response at
time $t$ to an impulsive perturbation applied at time 0) show an
exponential decay modulated in amplitude by oscillating
functions~\cite{FAZ09}.  Moreover violations of the EFDR $C(t)=R(t)$
(Einstein relation) are observed for $\alpha<1$~\cite{PBV07,VPV08}.

\begin{figure}[!htb]
\centering
\includegraphics[width=.7\columnwidth,clip=true]{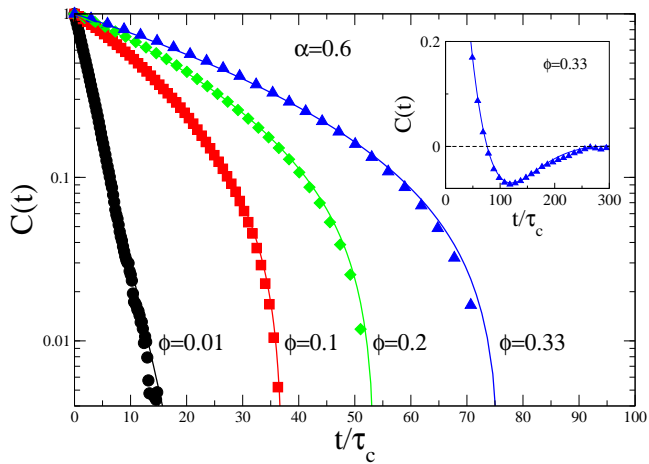}
\caption{(Color online). Semi-log plot of $C(t)$ (symbols) for
  different values of $\phi=0.01,0.1,0.2,0.33$ at $\alpha=0.6$. Times
  are rescaled by the mean collision time $\tau_c$.  The other
  parameters are fixed: $N=2500$, $m=1$, $M=25$, $r=0.005$, $R=0.025$,
  $T_b=1$, $\gamma_b=200$.  Times are rescaled by the mean collision
  times $\tau_c$, as measured in the different cases. Numerical data
  are averaged over $\sim 10^5$ realizations. Continuous lines are the
  best fits obtained with Eq.~(\ref{Cfit}). Inset: $C(t)$ and the best
  fit in linear scale for $\phi=0.33$ and $\alpha=0.6$.}
\label{fig_corr}
\end{figure}

Molecular dynamics simulations~\cite{SVGP10} of the system have been
performed, giving access to $C(t)$ and $R(t)$, for several different
values of the parameters $\alpha$ and $\phi$,  see Fig.~\ref{fig_corr}.

Notice that the Enskog approximation~\cite{BP04,SVCP10} cannot predict
the observed functional forms, because it only modifies by a constant
factor the collision frequency.  In order to describe the full
phenomenology, a model with more than one characteristic time is
needed.  The proposed model is a Langevin equation with a single
exponential memory kernel~\cite{BBR66,ZBCK05,SVGP10}, as in
Eq.~(\ref{langmemory})
\begin{equation} 
M\dot{ V}(t)=-\int_{-\infty}^tdt'~\Gamma(t-t') V(t')+ {\cal E}'(t),
\label{langmemory2}
\end{equation}
 in this case,
\begin{equation}
\Gamma(t)=2\gamma_0\delta(t)+\gamma_1/\tau_1e^{-t/\tau_1} 
\end{equation}
and ${\cal E}'(t)={\cal E}_0(t)+{\cal E}_1(t)$, with
\begin{equation}
\langle{\cal
  E}_0(t){\cal E}_0(t') \rangle=2 T_0\gamma_0\delta(t-t'),\\
\langle{\cal E}_1(t){\cal E}_1(t') \rangle=T_1\gamma_1/\tau_1e^{-(t-t')/\tau_1}
\end{equation}
 and $\langle {\cal E}_1(t){\cal E}_0(t') \rangle=0$.  In
the limit $\alpha\to 1$, the parameter $T_1$ is meant to tend to $T_0$
in order to fulfill the EFDR of the $2$nd kind $\langle {\cal
  E}'(t){\cal E}'(t')\rangle=T_0\Gamma(t-t')$. Within this model the
dilute case is recovered if $\gamma_1\to 0$.  In this limit, the
parameters $\gamma_0$ and $T_0$ coincide with $\Gamma_E$ and
$T_{tr}^E$ of the Enskog theory~\cite{SVCP10}.  The exponential form
of the memory kernel can be justified within the mode-coupling
approximation scheme, see~\cite{SVGP10} for details.

The model~(\ref{langmemory2}) predicts $C=f_C(t)$ and $R=f_R(t)$ with
\begin{equation}
f_{C(R)}=e^{-gt}[\cos(\omega t)+a_{C(R)}\sin(\omega t)]. \label{Cfit} 
\end{equation} 
The variables $g$, $\omega$, $a_C$ and $a_R$ are known algebraic
functions of $\gamma_0$, $T_0$, $\gamma_1$, $\tau_1$ and $T_1$. In
particular, the ratio $a_C/a_R=[T_0-\Omega (T_1-T_0)] /[T_0+\Omega
  (T_1-T_0)]$, with
$\Omega=\gamma_1/[(\gamma_0+\gamma_1)(\gamma_0/M\tau_1-1)]$.  Hence,
in the elastic ($T_1 \to T_0$) as well as in the dilute limit
($\gamma_1 \to 0$), one gets $a_C=a_R$ and recovers the EFDR
$C(t)=R(t)$. In Fig.~\ref{fig_corr} the continuous lines show the
result of the best fits obtained using Eq.~\eqref{Cfit} for the
correlation function, at restitution coefficient $\alpha=0.6$ and for
different values of the packing fraction $\phi$. The functional form
fits very well the numerical data.  A fit of measured $C$
and $R$ against Eqs.~(\ref{Cfit}), together with a measure of $\langle
V^2\rangle$, yields five independent equations to determine the five
parameters entering the model.  We used the external parameters
mentioned before, changing $\alpha$ or the box area $A$ (to change
$\phi$) or the intruder's radius $R$, in order to change $\phi$
keeping or not keeping constant $\gamma_g \sim 1/l_0 \to 0$ (indeed
different dilute limits can be obtained, where collisions matter or
not). Fits from numerical simulations suggest the following
identification for the parameters: $\gamma_0 \sim \Gamma_E$, $T_0 \sim
T_{tr}$ and $T_1 \sim T_b$. The coupling time $\tau_1$ increases with
the packing fraction and, weakly, with the inelasticity. In the most
dense cases it appears that $\gamma_1 \sim \gamma_g^E \propto \phi$:
such an observation however does not hold in the dilute limit at
constant collision rate, where $\gamma_1 \to 0 \ll \gamma_g^E$. It is
also interesting to notice that at high density $T_{tr} \sim T_g \sim
T_g^E$, which is probably due to the stronger correlations among
particles. Finally we notice that, at large $\phi$, $T_{tr}>T_{tr}^E$,
which is coherent with the idea that correlated collisions dissipate
{\em less} energy.

Looking for an insight of the relevant physical mechanisms underlying
such a phenomenology and in order to make clear the meaning of the
parameters, it is useful to map Eq.~(\ref{langmemory2}) onto a
Markovian equivalent model by introducing an auxiliary field
(see~\cite{VBPV09,SVGP10} for details on how the mapping is achieved):
\begin{eqnarray}
M\dot{V}&=&-\Gamma_E(V-U)+\sqrt{2\Gamma_E T_g}{\cal E}_V \nonumber \\
M'\dot{U}&=&-\Gamma' U - \Gamma_E V+\sqrt{2 \Gamma' T_b} {\cal E}_U,
\label{local_field}
\end{eqnarray}
where ${\cal E}_V$ and ${\cal E}_U$ are indipendent white noises of unitary
variance and we have exploited the numerical observations discussed above (i.e. $T_0 \sim T_g$, $T_1 \sim T_b$ and $\gamma_0 \sim \Gamma_E$), while $\Gamma'=\frac{\gamma_0^2}{\gamma_1}$ and $M'=\frac{\gamma_0^2\tau_1}{\gamma_1}$.
In the chosen form~(\ref{local_field}), the dynamics of the tracer is
remarkably simple: indeed $V$ follows a memoryless Langevin equation
in a {\em Lagrangian frame} with respect to a local field $U$, which
is the \emph{local average velocity field} of the gas particles
colliding with the tracer. Extrapolating such an identification to
higher densities, we are able to understand the value for most of the
parameters of the model: the self drag coefficient of the intruder in
principle is not affected by the change of reference to the Lagrangian
frame, so that $\gamma_0 \sim \Gamma_E$; for the same reason $T_0 \sim
T_{tr}$ is roughly the temperature of the tracer; $\tau_1$ is the main
relaxation time of the average velocity field $U$ around the Brownian
particle; $\gamma_1$ is the intensity of coupling felt by the
surrounding particles after collisions with the intruder; finally
$T_1$ is the ``temperature'' of the local field $U$, easily identified
with the bath temperature $T_1 \sim T_b$: indeed, thanks to momentum
conservation, inelasticity does not affect the average velocity of a
group of particles which almost only collide with themselves.

\subsection{Generalized FDR and entropy production}
\label{Generalized FDR and entropy production}

\begin{figure}[!htb]
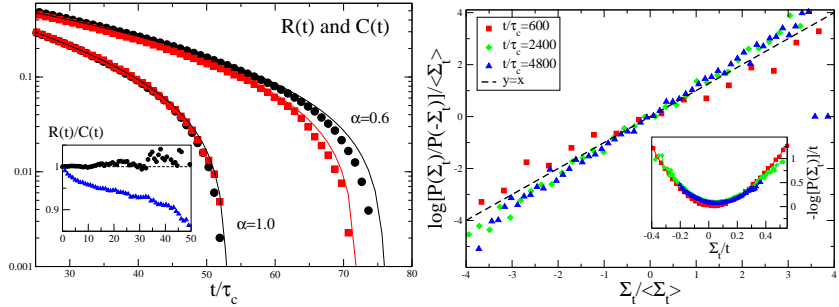

\includegraphics[width=.45\columnwidth,clip=true]{tracer2.eps}
\includegraphics[width=.45\columnwidth,clip=true]{tracer3.eps}
\caption{(Color online). Left: correlation function $C(t)$ (black circles) and response function $R(t)$ (red
squares) for $\alpha=1$ and $\alpha=0.6$, at $\phi=0.33$. 
Continuous lines show the best fits curves obtained with Eqs.~(\ref{Cfit}). 
Inset: the ratio $R(t)/C(t)$ is reported in the same cases. Right: Check of the fluctuation
relation~(\ref{GCrel}) in the system with $\alpha=0.6$ and
$\phi=0.33$. 
Inset: collapse of the rescaled probability
distributions of $\Sigma_t$ at large times onto the large deviation function.}
\label{fig_resp}
\end{figure}

The model discussed above, Eq.~(\ref{local_field}), is able to
reproduce the violations of EFDR, as show in Fig.~\ref{fig_resp},
which depicts correlation and response functions in a dense case
(elastic and inelastic). In the inelastic case, deviations from EFDR
$R(t)=C(t)$ are observed. In the inset of Fig.~\ref{fig_resp} the
ratio $R(t)/C(t)$ is also reported.  As shown in
Section~\ref{Generalized Fluctuation-Dissipation relations}, a
relation between the response and correlations measured in the
unperturbed system still exists, but - in the non-equilibrium case -
must take into account the contribution of the cross correlation
$\langle V(t)U(0) \rangle$, i.e.:
\begin{equation}
R(t)=a C(t) + b \langle V(t)U(0)\rangle \label{totresp}
\end{equation}
with $a=[1-\gamma_1/M(T_0-T_1)\Omega_a]$ and $b=(T_0-T_1)\Omega_b$,
where $\Omega_a$ and $\Omega_b$ are known functions of the parameters. At 
equilibrium, where $T_0=T_1$, the EFDR is recovered.

An important independent assessment of the effectiveness of
model in Eq.~\eqref{local_field}  comes from the study of the fluctuating
entropy production~\cite{seifert05} which quantifies the deviation
from detailed balance in a trajectory. Given the trajectory in the
time interval $[0,t]$, $\{V(s)\}_0^t$, and its time-reversed $\{{\cal
  I}V(s)\}_0^t\equiv\{-V(t-s)\}_0^t$, as discussed in
section \ref{entropy_production}, shown that the entropy production
for the model~(\ref{local_field}) takes the form~\cite{PV09}
\begin{equation}
\Sigma_t=\log\frac{P(\{V(s)\}_0^t)}{P(\{{\cal I}V(s)\}_0^t)}
\approx \gamma_0\left(\frac{1}{T_0}-\frac{1}{T_1}\right)\int_0^t ds~V(s)U(s).
\label{entropy_prod}
\end{equation}
This functional vanishes exactly in the
elastic case, $\alpha=1$, where equipartition holds, $T_1=T_0$, and is
zero on average in the dilute limit, where $\langle VU\rangle=0$.

Formula~\eqref{entropy_prod} reveals that the leading source of
entropy production is the energy transferred by the ``force''
$\gamma_0 U$ on the tracer, weighed by the difference between the
inverse temperatures of the two ``thermostats''. Therefore, to measure
entropy production, we need to measure the fluctuations of $U$: from
the above discussion we have for $U$ the interpretation of a
spontaneous local velocity field interacting with the tracer.
Therefore it can be measured doing a local average of particles'
velocity in a circle of radius $l$ centered on the
tracer. Details on how to choose in a reliable way the proper $l$ are
given in~\cite{SVGP10}, for instance following such procedure, in the case
$\phi=0.33$ and $\alpha=0.6$, we estimate for the
correlation length $l \sim 9r \sim 6l_0$. Then, measuring 
 the entropy production from Eq.~(\ref{entropy_prod}) 
along many trajectories of length $t$, we can
 compute the probability $P(\Sigma_t=x)$ and compare it to
$P(\Sigma_t=-x)$, in order to verify the Fluctuation Relation
\begin{equation}
\log\frac{P(\Sigma_t=x)}{P(\Sigma_t=-x)}=x.
\label{GCrel}
\end{equation}
In the right frame of Fig.~\ref{fig_resp} we report our numerical results. The main frame
confirms that at large times the Fluctuation Relation~(\ref{GCrel}) is
well verified within the statistical errors.  The inset shows the
collapse, for large times, of $\log P(\Sigma_t)/t$ onto the Cramer
function $S(y=\frac{\Sigma_{t}}{t})$, introduced in Section~\ref{Fluctuation Theorem}.  Notice also that formula~\eqref{entropy_prod}
does not contain further parameters but the ones already determined by
correlation and response measure, i.e. the slope of the graph is not
adjusted by further fits. Indeed a wrong evaluation of the weighing
factor $(1/T_0-1/T_1) \approx(1/T_{tr}-1/T_b)$ or of the ``energy
injection rate'' $\gamma_0 U(t) V(t)$ in Eq.~\eqref{entropy_prod}
could produce a completely different slope in Fig.~\ref{fig_resp} (right frame).

To conclude this section, we stress that velocity correlations
$\langle V(t) U(t') \rangle$ between the intruder and the surrounding
velocity field are responsible for both the violations of the EFDR and
the appearance of a non-zero entropy production, provided that the two
fields are {\em at different temperatures}.  We also mention that
larger violations of EFDR can be observed using an intruder with a
mass equal or similar to that of other particles~\cite{PBV07}, with
the important difference that in such a case a simple
``Langevin-like'' model for the intruder's dynamics is not available.



\section{Conclusions and perspectives}

We have reviewed a series of recent results on
fluctuation-dissipation relations for out-of-equilibrium systems. The {\em leitmotiv} of
our discussion is the importance of correlations for non-equilibrium
response, much more relevant than in equilibrium systems: indeed the
Generalized Fluctuation-Dissipation relation discussed in
Section~\ref{Comb model}, \ref{Langevin process}, \ref{GranularIntruder} deviates
from its equilibrium counterpart for the appearance of additional
contributions coming from correlated degrees of freedom. This is the
case, for instance, of the linear response in the diffusion model on a
comb-lattice, where - in the presence of a net drift - the linear
response takes a non-negligible additive contribution, see
Eq.~\eqref{ale1}. The same occurs in the
general two-variable Langevin model: additional contribution to the
equilibrium linear response appears when the main field $V$ is coupled
to the second field $U$, which only appears out-of-equilibrium, see
Eq.~\eqref{GFRlangevin}. The granular intruder is a realistic
many-body instance of such a coupling scenario, where we have seen the strong
influence of correlated degrees of freedom to the linear response, see
Eq.~\eqref{local_field}.

Remarkably, in some cases, one may explicitly verify that the coupled
field which contributes to the linear response in non-equilibrium
setups is also involved in the violation of detailed balance: such a
violation is measured by the fluctuating entropy production. The
connection between non-equilibrium couplings and entropy production has
been discussed for general coupled linear Langevin models, see
Eq.~\eqref{EntropyTracer} and then verified for the entropy production of the
granular intruder, Eq.~\eqref{entropy_prod}.

In conclusion many results point in the same direction, suggesting a
general framework for linear-response in systems with non-zero entropy
production. Even in out-of-equilibrium configurations, a clear
connection between response and correlation in the unperturbed systems
exists. A further step is looking for more accessible observables
which could make easier the prediction of linear response: indeed,
both proposed formula, Eq.~\eqref{3.9} and Eq.~\eqref{2.9}, require
the measurement of variables which, in general, depend on full
phase-space (microscopic) observables and can be strictly
model-dependent. Such a difficulty also explains why, in a particular
class of slowly relaxing systems with several well separated
time-scales, such as spin or structural glasses in the aging dynamics,
i.e. after a sudden quench below some dynamical transition
temperature, approaches involving ``effective temperatures'' have been
used in a more satisfactory way~\cite{CR03}. More recent
interpretations of the additional non-equilibrium contributions has
been proposed in~\cite{BMW09}, but its predictive power is
not yet fully investigated and represents an interesting line of
ongoing research.

\subsection*{Acknowledgments}

We thank A. Baldassarri, F. Corberi, M. Falcioni, E. Lippiello, U. Marini Bettolo Marconi,
L. Rondoni and M. Zannetti for a long collaboration on the issues here discussed.

\bibliography{fluct.bib}
\bibliographystyle{unsrt}

\end{document}